\documentclass[runningheads,anonymous]{llncs}
\usepackage[T1]{fontenc}
\usepackage{graphicx}
\usepackage[rightcaption]{sidecap}
\usepackage{mathpartir}
\usepackage{xcolor}
\usepackage{amsmath}
\usepackage{orcidlink}

\newcommand{\arancione}[1]{\textcolor{orange}{#1}}

\newcommand{\rosso}[1]{\textcolor{red}{#1}}
\newcommand{\verde}[1]{\textcolor{teal}{#1}}

\newcommand{\location}[3]{\left\langle #1, #2, #3\right\rangle}
\newcommand{\para}{\;|\;}
\newcommand{\exec}[1]{\mathtt{exec}(#1)}
\newcommand{\transfer}[1]{\mathtt{tran}(#1)}
\newcommand{\transrec}[1]{\underline{\mathtt{tran}}(#1)}
\newcommand{\rec}[1]{\mathtt{rec}(#1)}
\newcommand{\lts}[1]{\xrightarrow{\; #1 \;}}

\newcommand{\conf}[1]{{\bf{conf}}(#1)}
\newcommand{\confs}{{\bf{conf}}}
\newcommand{\confrecs}{\underline{{\bf{conf}}}}
\newcommand{\confrec}[1]{\underline{{\bf{conf}}}(#1)}
\newcommand{\init}[1]{\mathtt{init}(#1)}

\newcommand{\mapping}{\mathcal{M}}
\newcommand{\nul}{\mathbf{0}}
\newcommand{\Data}[1]{D_{#1}}
\newcommand{\trace}{t}
\newcommand{\workflow}{\mathtt{W}}
\newcommand{\pointer}{^{\triangleright}}
\newcommand{\Out}{\mathit{O}}
\newcommand{\In}{\mathit{I}}
\newcommand{\msg}[1]{m_{#1}}
\newcommand{\error}[1]{\mathit{err}(#1)}
\newcommand{\context}[1]{\mathbf{\;T[\;}#1\mathbf{\;]\;}}
\newcommand{\ok}[1]{ok_{#1}}

\begin{document}

\title{A Fault Tolerance Mechanism for Hybrid Scientific
Workflows}

\author{Alberto Mulone\inst{1}\orcidlink{0009-0009-2600-613X} \and
Doriana Medić\inst{1}\orcidlink{0000-0002-7163-5375} \and
Marco Aldinucci\inst{1}\orcidlink{0000-0001-8788-0829}}
\authorrunning{A. Mulone, D Medić and M. Aldinucci}
\institute{University of Turin, Italy,
\email{\{name.surname\}@unito.it}}
\maketitle              % typeset the header of the contribution
\begin{abstract}
In large distributed systems, failures are a daily event occurring frequently, especially with growing numbers of computation tasks and locations on which they are deployed.
The advantage of representing an application with a workflow is the possibility of exploiting Workflow Management System~(WMS) features such as portability. A relevant feature that some WMSs supply is reliability.
Over recent years, the emergence of hybrid workflows has posed new and intriguing challenges by increasing the possibility of distributing computations involving heterogeneous and independent environments. Consequently, the number of possible points of failure in the execution increased, creating different important challenges that are interesting to study.
This paper presents the implementation of a fault tolerance mechanism for hybrid workflows based on the recovery and rollback approach. A representation of the hybrid workflows with the formal framework is provided, together with the experiments demonstrating the functionality of implementing approach.
\end{abstract}

\section{Introduction}
 
In contemporary contexts, the interest in creating and deploying intricate applications across expansive networks of diverse computing architectures is increasing in different scientific domains. Many Workflow Management Systems~(WMS) are developed to cope with the demands of applications coming from various fields, from geophysics~\cite{Klampanos2020} to bioinformatics~\cite{mulone_porting_2023}, simulation of chemical reactions~\cite{COLONNELLI2022282} and astrophysics~\cite{SCIACCA2015146}. 
The necessity to employ different locations to execute a computation has different reasons, such as computational performance or data privacy.
The heterogeneity of computing architectures is an increasing trend because new specialised hardware is being developed to execute some tasks performatively. % todo: add reference?
Moreover, data are the most important asset today, and nobody wants to share them for many different reasons, for example data can be sensitive and they are protected for legal reasons. Thus, sometimes, it is impossible to move data, and it is necessary to use them carefully.

Hybrid workflow~\cite{FERREIRADASILVA2017228} systems are a possible solution for these requirements in the workflow application. They can span the steps in multiple and heterogeneous environments without sharing memory zones.
As said before, hybrid WMSs can move the data to execute a step in a more suitable location based on the step requirements, such as a specific processor, or they can move the computation to minimise the data movements.
For example, hybrid workflow allows the modelling of federated applications, where it cannot be possible to move the data, e.g. orchestrating a federated Deep Neural Network training across different HPC centres~\cite{22:ml4astro}.
Some of WMS implementing this paradigm are DagOnStar~\cite{dagonstar}, Pegasus~\cite{pegasus:2019} and StreamFlow~\cite{20Lstreamflow:tetc}.

When workflow applications run on large distributed systems, failure is not a possibility but a daily event whose occurrence increases with the number of activities and actors involved. 
WMS execution depends on many layers, often perceived as black boxes. It is not guaranteed that the application used to compute steps or the locations used to deploy steps are fault tolerant. 
Hence, implementing a fault tolerance mechanism directly in WMS to recover the failed step execution is becoming inevitable. 

The contribution of this paper is two-sided, on one side providing the implementation of the fault tolerance mechanism for hybrid workflow systems (Sec.~\ref{sec:implementation}) based on the recovery and rollback techniques (detailed description given in Sec.~\ref{sec:fault-tolerance-approach}), and, on the other side, giving the formalisation of the behaviour of the implemented approach (Sec.~\ref{subsec:formalisation}). 
The formal description of the actions of a hybrid workflow is given through an example (Sec.~\ref{sec:formal-example}).
The fault tolerance mechanism is implemented on StreamFlow, and an experiment with a workflow execution containing the simulated fault behaviour is provided (Sec.~\ref{sec:experiments}).

\section{Related work} \label{sec:related-works}

Different types of errors can occur during the workflow execution. They are soft, fail-stop and silent errors~\cite{henri-colleagues}. The soft or fail-stop errors cause the failure of the step execution. When a soft error occurs, the input data of the step is still undamaged in the executing locations. Instead, the data is corrupted or lost when a fail-stop error occurs. Finally, detecting the silent errors is more complex since the step terminates with success, but the output data is contaminated (i.e. the step produces the wrong result).

In the literature, various works deal with the mechanisms for error management involving other fields beyond the scientific workflows. Mainly, these mechanisms can be divided into two categories, following the notation of \cite{ft-survey}, the one dealing with the error when the failure occurs, so-called \emph{reactive methods}, or the one trying to predict and prevent the error, so-called \emph{proactive methods}. Both approaches require additional work to guarantee reliability, such as saving metadata at runtime, duplicating data, increasing the execution time or the number of resources needed compared to an unreliable execution. Therefore, the choice of the method utilised is guided by the type of failure to manage and the availability of resources (time and funding). This paper focuses on reactive methods and different techniques used to recover the failure.

A simple fault tolerance approach is \emph{retry-rollback}, which re-executes the failed step. However, the step can have some dependencies on other steps, which obliges their rollback~\cite{rollback-message-passing}. In the worst scenario, the workflow is fully re-executed, describing the domino effect~\cite{randell}.
This approach, usually, is coupled with \emph{checkpoint} in which the WMS saves the workflow state and the data at runtime by applying different heuristics~\cite{checkpoint-henri}. The recovery of the failure is modelled by bringing the state of the workflow back to the last checkpoint (its valid state) and restarting the execution from there.

Another technique is \emph{replication}, where the WMS executes the step with the same input data multiple times in parallel; hence, if one replica fails, other replicas are still alive, ensuring that the workflow executions continue. This mechanism can also be used to detect silent errors~\cite{replica-and-silent-error}.

% rescue dag
Finally, the \emph{rescue DAG} approach, implemented in DAGMan~\cite{dagman}, saves metadata when a failure occurs and continues to run the workflow until it is possible. In the state where the workflow has only failed and pending steps (waiting for the output of the failed steps), it terminates the execution and creates a new workflow (using saved metadata) containing only the missing steps. This new workflow continues the execution of the original workflow.

\section{The fault tolerance mechanism for hybrid workflows} \label{sec:formalization}

This section describes the considered workflow setting and the fault tolerance implemented, giving a formal syntactical representation of the hybrid workflow and its fault-tolerant mechanism.
The workflow is represented as Direct Cyclic Graph (DCG), where the vertices are the steps to be computed and the edges are the dependencies between the steps given by data dependency. It is supposed that the step execution is deterministic since many of the workflow applications we are dealing with are deterministic. Moreover, the step is seen as a black box; thus, the step re-execution is done from the beginning, regardless of its internal state when the failure occurs. The WMS discards the data produced by a failed step because they can be incomplete. Finally, the step can not operate the data in-place, thus input data are still valid if they exist after the failure.
The hybrid workflow is obtained by mapping steps to different locations, assuming that the locations are mutually connected following some topology\cite{colonnelli_2022_7273357}.

\subsection{The implemented fault tolerance approach}\label{sec:fault-tolerance-approach}

The fault tolerance mechanism employed in this paper is based on the \emph{recovery workflow} approach. Creating a separate workflow aims to retrieve the valid state of the failed steps while the non-affected continue to run their executions. After recovering the failed step, its output is returned to the original workflow. The support of the loops and the concurrent executions between the original workflow and the recovery workflows are the main differences between our solution and the DAGMan approach. The existence of two types of failure-free entities is assumed: WMS, in this paper called \emph{driver} or \emph{driver location}, and all locations on which the initial dataset is stored. In this way, it is guaranteed that the initial input to re-execute the whole workflow (if necessary) will not be lost.

The presented implementation manages two types of errors, soft and fail-stop errors, that occur during step execution. Specifically, during the implementation, the considered failures are the application failure (e.g., a segmentation fault occurs) and the location failure (e.g., the HPC system has extraordinary maintenance while a step is running). 
In the case of application failure, it is enough to re-execute the step on the location\footnote{It is possible also to change location, but other policies are necessary, such as copying the input data to a new location if it is needed}.
Instead, in the case of location failure, all the data on the location can be lost when it has ephemeral memory (e.g., in the Kubernetes Pods without a permanent volume). In this case, there are two possible scenarios. It is possible to copy the input data to the failed location, and the same strategy of application failure can be applied if the data are available in another location. Otherwise, if the data is not present in any location, it is necessary to roll back and recover the steps that produced the lost data.

As mentioned in Sec.~\ref{sec:related-works}, the fault tolerance mechanism introduces some overhead. In this approach, the overhead is given by metadata collection and increased workflow execution time. 
The metadata tracks the step data produced, the required inputs, and the location where the data resides. These metadata are stored by the driver; however, the data are across different locations. Thus, the driver saves a token that represents the data. 
They are collected as a graph, called \emph{provenance graph}, where the vertices are the data information and the edges are the dependencies between the data.
In the implementation, the steps to rollback are decided through the bottom-up navigation of the provenance graph with a Breadth-First Search (BFS) visit. The starting data of the navigation are the input data of the failed step. 
If the data of the provenance graph vertex is available in some locations, it becomes an input of the recovery workflow.
Otherwise, it is necessary to visit the parent vertices of the current vertex.

\subsection{Syntactical representation}\label{subsec:formalisation}

This section gives a hint on a formalisation of the fault tolerance mechanism for the hybrid workflows. The workflow description follows the idea presented in~\cite{colonnelli_2022_7273357}, while the formalisation is inspired by the distributed $\pi$-calculus approach~\cite{distributedPi} to model location aware workflows~\cite{medic-wide23} with a more elaborated framework to be able to cover the recovery part of the workflow.
As in~\cite{colonnelli_2022_7273357}, the information about the location is recorded into the \emph{location configuration}, in this case, containing the name of the location~$l$, the set of the data and messages saved at the location~$l$ (denoted by $\Data{l}$) and the trace~$\trace$ of the actions to be performed on the location~$l$. Another addition respect to the semantics given in~\cite{colonnelli_2022_7273357} is the modelling of the \emph{driver location}, which contains the additional elements, saved into the set $\Data{l_d}$, the trace of the whole workflow structure $\trace_{\workflow}$ and the mapping function $\mapping(s)$ that maps the step $s$ to the location on which it is deployed. The driver location has a global knowledge of the workflow and its execution. It orchestrates the execution of the steps: knows when the data is produced or received, manages all recovery actions, and so on.

The traces are built by the following actions: \emph{execution} of the step $s$, denoted as $\exec{s,\In,\Out}$ recording sets of input and output data of the step $s$;
\emph{transfer}, denoted by couple of prefixes $\transfer{v,l_2}$ and $\transrec{v,l_1}$ modeling the transfer of the value $v$ from location $l_1$ to the location $l_2$. The value $v$ could be the data $d$ or different messages: $\msg{d,l}$, sent from location $l$ to the driver, signalling that the data $d$ is produced on the location $l$; $\msg{s}$, sent from the driver to indicate that the step $s$ can be executed; $\ok{d}$ and $\ok{\msg{s}}$, sent to the driver to confirm the reception of the corresponding data/message; $\error{s}$, sent to the driver location indicating that the step $s$ failed (soft error) and the data to re-execute it are still present on the location; $\error{D,l}$, transferred to the driver indicating that the data contained in the set $D$ is missing on the location $l$ due to the location fails (fail-stop error); $\error{d,l}$ and $\error{\msg{s}}$ indicating that the data and the message are not received on the location $l$ due to the transfer failure.
Finally, $\rec{x}$ is the recovery of the failed element $x$. This element can be the step $s$, in which the application fails, the lost data, or the value not received due to transfer failure.

The basic actions are composed in the trace by applying different operators, sequential execution $.$, parallel composition $\para$ and the choice operator $+$. Differently from the classic process algebra, the prefix representing the executed step is not discarded but annotated with the pointer $\pointer$. For instance, considering the trace $\trace=\exec{s,\In,\Out}.\transfer{v,l_2}$, in the classical case, after the execution of the step, the obtained trace is $\trace'=\transfer{v,l_2}$, representing only the actions to be executed in the future. In our case, the initial trace $\trace$ is annotated with the pointer, indicating the state of the execution; therefore, the trace $\trace$ becomes $\pointer\exec{s,\In,\Out}.\transfer{v,l_2}$, and after the execution of the step, the obtained trace is $\trace''=\exec{s,\In,\Out}.\pointer\transfer{v,l_2}$ indicating that the next action to be performed is transfer of the value $v$. 
Formally:
\begin{definition} \label{def:syntax}
      The grammar defines the syntax of a workflow system $\workflow$:
      \vspace{-1mm}
      {\footnotesize
       \begin{align*}
         \workflow &::= \location{l}{\Data{l}}{\trace}\parallel
                     (\workflow_1 \para \workflow_2)\\
         \trace &::= \mu \parallel 
                    \trace_1.\trace_2 \parallel 
                    (\trace_1\para\trace_2)\parallel
                    (\trace_1+\trace_2)\parallel 
                    \pointer\trace \parallel
                    \nul\\
         \mu &::= \exec{s,\In,\Out} \parallel
                    \transfer{v,l_2}\parallel \transrec{v,l_1}\parallel 
                   \rec{x}\\
          v & ::= d \parallel
                 \msg{s}\parallel\msg{d,l}\parallel \ok{\msg{s}}\parallel \ok{d}\parallel
                 \error{x}\\
        x &::= s\parallel D,l \parallel d,l \parallel \msg{s},l
       \end{align*}}
    \end{definition}

The confirmation of the data/message received can be modelled using the syntax of Def.~\ref{def:syntax} in the following way:
\[
\conf{x,l}=\transfer{\ok{x},l_d}+\transfer{\error{x,l},l_d} \text{ where } x=\msg{s} \;\vee\; x=d.
\]
For instance, location $l$ confirms to the driver location the reception of the data $d$ (denoted as $\conf{x,l}$) by transferring to it or the message $\ok{d}$, meaning that the data is received successfully, or $\error{d,l}$ indicating that the data is not received on the location~$l$.

{\bf{Notation.}} We introduce some simplifications of the notation, when possible, to improve the paper's readability. Therefore, we write $\exec{s}$, instead of $\exec{s, \In,\Out}$ and $\transrec{x,l}.\confs$ instead of $\transrec{x,l}.\conf{\ok{x},l'}$ since from $\transrec{x,l}$ is clear what is the message which reception is necessary to confirm.
To simplify the traces of the step (actions necessary to execute the step and transfer the data to the succeeding computations), the trace regarding the step~$s$ can be modelled as: input actions, divided in input of the message $\msg{s}$ and the input of necessary data (denoted as trace $\trace_{\In(s)} = \prod_{j}\transrec{d_j,l_h}.\conf{\ok{d_j},l_h}$); execution of the step; actions signalling to the driver about the outcome of the step (confirming the produced data or indicating an error in the execution); and output actions transferring data to the targeting locations ($\trace_{\Out(s)}$). Considering this, the trace of any step $s$ can be represented with:
{\footnotesize
\begin{align*}
\trace(s)= &(\transrec{\msg{s},l_d}.\confs\para \trace_{\In(s)}).\exec{s}.(\prod_{i} (\transfer{\msg{d_i,l},l_d} + \transfer{\error{x},l_d})).\trace_{Out(s)}
\end{align*}}
To simplify, the notation $\trace(s_{\setminus \trace_{\In(s)}})$ denotes the trace $\trace(s)$ without the input data part ($\trace_{\In(s)}$). 
The trace on the driver location which represents the orchestration of the step $s$ mapped to the location $l$ is complementary to the $\trace(s)$ and it is modelled as:
{\footnotesize
\begin{align*}
\trace_d(s)=  (\transfer{\msg{s},l}.\confrecs\para \trace_{d_{\In(s)}}).(\prod_{i} (\transrec{\msg{d_i,l},l} + \transrec{\error{x},l}))
\end{align*}}
where $\trace_{d_{\In(s)}}=\prod_{j}\transfer{d_j,l}.\confs\para \prod_{h} \conf{\ok{d_h},l} $ means that the driver sends data~$d_j$ to the step $s$ and receives the confirmations~$\ok{d_h}$ of the data received on location~$l$.

\section{Formalisation through the example}\label{sec:formal-example}

\begin{SCfigure}[][t]
    \centering
    \includegraphics[width=.38\linewidth]{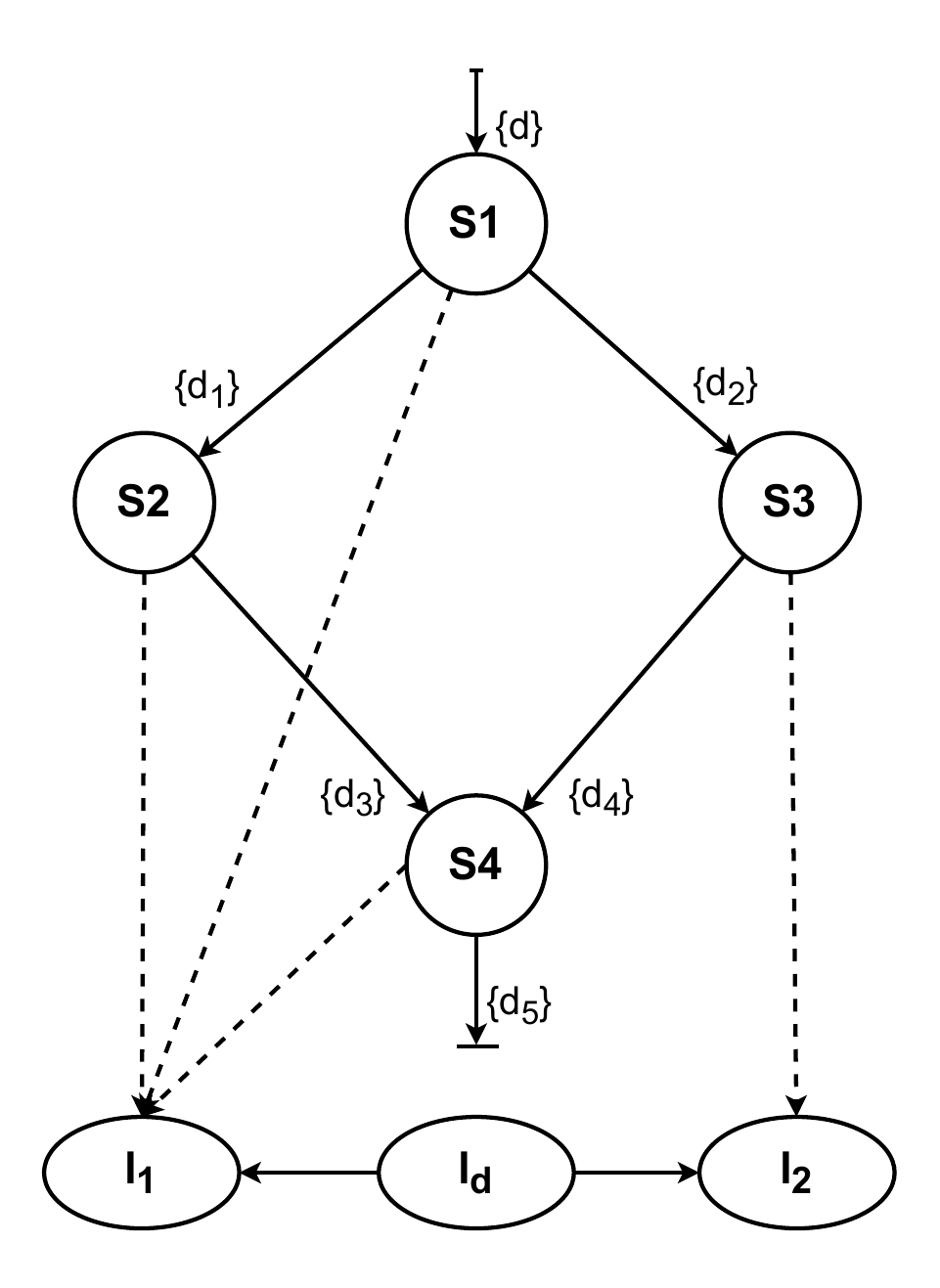}%
    \caption{
        \small{The hybrid workflow consisting of four steps (represented as circles), connected in the following way: step $s_1$ produces data $d_1$ and $d_2$ necessary for the execution of the steps $s_2$ and $s_3$, respectively. The resulting data of those two steps trigger the execution of the step $s_4$. The steps $s_1, s_2$ and $s_4$ are deployed on the location $l_1$, while step $s_3$ is mapped on the location $l_2$. The driver location contains the initial input data of the workflow, and it orchestrates the executions.}
    }
    \label{fig:example}
    \vspace{-2mm}
    \end{SCfigure}

Representation of the hybrid workflows in the formal language defined with Def.~\ref{def:syntax} is given by modelling the workflow provided in Fig.~\ref{fig:example}. For the sake of paper readability, we show the mechanism by concentrating on the specific action, while the semantics rules (Figs.~\ref{fig:semantics} and \ref{fig:roll-semantics}), extensive definition of the workflow behaviour and illustration of the recovery mechanism are provided in App.~\ref{sec:appendix}.
It is assumed that the necessary contextual rules, together with the structural congruence rules like commutativity of the parallel and choice operator, neutral element for the sequence and parallel operator (element $\nul$), and so on, are holding.

The initial state of the workflow in Fig.~\ref{fig:example} can be represented using the syntax of Def.~\ref{def:syntax} in the following way:
{\footnotesize
\vspace{-2mm}
\begin{align*}
\workflow = \location{l_d}{\Data{l_d}}{\pointer\trace_d, \trace_{\workflow}}
        \para \prod_{i=1}^{4} \location{l_i}{\emptyset}{\emptyset}
\end{align*}}
where $\Data{l_d}=\{\trace_{\workflow},\mapping(s_i),d,\msg{s_i}\}$, $i=1,\ldots,4$ and $\trace_{\workflow}=\prod_{i=1}^{4} \trace(s_i)$. Trace $\trace_d$ consists of four subtraces $\trace_d(s_i)$, executing on the driver location and orchestrating the executions of the four steps $s_i$. The trace $\trace_{\workflow}$ comprises the four traces $\trace(s_i)$ representing the actions to perform on the locations on which the steps $s_i$ are deployed.  
For instance, $\trace (s_3)$ denotes the inputs of the data $d_2$ and message $\msg{s_3}$; the confirmation of the reception, the execution of the step which produces the new data $d_4$ or complementing error, then communicating to the driver about the value produced and when the data is produced, finally, transferring the data to the next step/location ($s_4/l_1$). Formally:
{\footnotesize
\begin{align*}
\trace(s_3)=& (\transrec{d_{2},l_1}.\conf{\ok{d_2},l_2}\para \transrec{\msg{s_{3}},l_d}.\confs).\\
&\exec{s_3}.(\transfer{\msg{d_{4},l_{2}},l_d}+\transfer{\error{x},l_d}).\transfer{d_{4},l_1}
\end{align*}
}
The corresponding trace on the driver location $\trace_d(s_3)$ would be:
{\footnotesize
\begin{align*}
\trace_d (s_3)=&\init{\trace(s_3),l_2}.(\transfer{\msg{s_3},l_2}. \confrecs\para\\
&\qquad\confrec{\ok{d_2},l_2}). (\transrec{\msg{d_4,l_2},l_2}+\transrec{\error{x},l_2})
\end{align*}
}
The driver first initialises the trace of the step on the deploying location, then receives the confirmation that necessary data is received on the location and triggers the step execution. At the end, it receives the produced data or the obtained error. A full representation of the traces can be found in App.~\ref{sec:appendix}.

The following text shows how some actions work, by using the traces $\trace(s_3)$ and $\trace_d (s_3)$. In particular are shown the transfer of the data, successful step execution and unsuccessful execution with recovery mechanism. 
To simplify the presentation and highlight the executing action, $\context{}$ denotes the context of the executing action. 
Therefore, the traces $\trace(s_3)$ and $\trace_d (s_3)$, which show the transfer of the message $\msg{s}$ from the driver location to the location $l_2$, can be written as:
{\footnotesize
\begin{align*}
\trace (s_3)=\context{\arancione{\pointer}\transrec{\msg{s_3},l_d}}\qquad\text{ and }\qquad
\trace_d (s_3)=\mathbf{T'}[\arancione{\pointer}\transfer{\msg{s_3},l_2}]
\end{align*}
}
Componing the traces and the locations, there is:
{\footnotesize
\begin{align*}
&\location{l_d}{\Data{l_d}}{\mathbf{T}[\arancione{\pointer}\transfer{\msg{s_3},l_2}]}   \para 
\location{l_2}{\Data{l_2}}{\mathbf{T'}[\arancione{\pointer}\transrec{\msg{s_3},l_d}]}\arancione{\lts{}}\\
%%%%%%%
&\location{l_d}{\Data{l_d}}{\mathbf{T}[\transfer{\msg{s_3},l_2}\arancione{\pointer}]}   \para 
\location{l_2}{\Data{l_2}\arancione{\cup \{\msg{s_3},\ok{\msg{s_3}}\}}}{\mathbf{T'}[\transrec{\msg{s_3},l_d}\arancione{\pointer}]}
\end{align*}
}

When all necessary elements are on the location $l_2$, the step $s_3$ can be executed successfully:
{\footnotesize
\begin{align*}
&\location{l_2}{\Data{l_2}}{\mathbf{T''}[\arancione{\pointer\exec{s_3}}]}\arancione{\lts{}}\location{l_2}{(\Data{l_2}\setminus \msg{s_3})\verde{\cup\{\msg{d_4,l_2}, d_4\}}}{\mathbf{T''}[\exec{s_3}\arancione{\pointer}]}
\end{align*}
}

Considering the failure of a step, for instance, the computation of the step has failed and the data is still present on the location $l_2$, then the computation would be:
{\footnotesize
\begin{align*}
&\location{l_2}{\Data{l_2}}{\mathbf{T''}[\arancione{\pointer\exec{s_3}}]}\arancione{\lts{}}
\location{l_2}{(\Data{l_2}\setminus \msg{s_3})\rosso{\cup\{\error{s_3}\}}}{\mathbf{T''}[\exec{s_3}\arancione{\pointer}]}
\end{align*}
}

To recover the error, the failure is communicated to the driver location by adding the $\rec{s_3}$ trace to the driver execution trace:
{\footnotesize
\begin{align*}
&\location{l_2}{\Data{l_2}}{\mathbf{T'''}[\arancione{\pointer}\transfer{\error{s_3},l_d}]}\para 
\location{l_d}{\Data{l_d}}{\mathbf{T^{iv}}[\arancione{\pointer}\transrec{\error{s_3},l_2}]} \lts{}\\
& \location{l_2}{\Data{l_2}\setminus \error{s_3}}{\mathbf{T'''}[\transfer{\error{s_3},l_d}\arancione{\pointer}]}\para\location{l_d}{\Data{l_d}\cup \error{s_3}}{\mathbf{T^{iv}}[\transrec{\error{s_3},l_2}\arancione{\pointer}]\rosso{\para \pointer\rec{s_3}}}
\end{align*}
}

By executing the recover action, the driver location is coordinating the necessary computation by repeating the step execution without transferring the data since it is already present on the location:
{\footnotesize
\begin{align*}
&\location{l_d}{\Data{l_d}}{\trace'_d\rosso{\para \pointer\rec{s_3}}}\para
\location{l_2}{\Data{l_2}}{\trace'(s_3)}\lts{}\\
%%%%%%%%%
&\location{l_d}{\Data{l_d}\setminus \error{s_3}}{\trace'_d\para \rec{s_3}\rosso{\pointer}\para\verde{\pointer\trace_d(s\setminus \trace_{d_{\In(s_3)}})}}
\location{l_2}{\Data{l_2}}{\verde{\nul\para \pointer\trace(s_3\setminus \trace_{\In(s_3)} ) }}
\end{align*}
}
where $\trace(s_3\setminus \trace_{\In(s_3)} )$ (similar for the trace $\trace_d(s\setminus \trace_{d_{\In(s_3)}})$) is:
{\footnotesize
\begin{align*}
\trace(s_3\setminus \trace_{\In(s_3)} )=& \transrec{\msg{s_{3}},l_d}.\confs.\exec{s_3}.\\
&(\transfer{\msg{d_{4},l_{2}},l_d}+\transfer{\error{x},l_d}).\transfer{d_{4},l_1}
\end{align*}
}

\section{Implementation and experiments}\label{sec:experiments}

\subsection{Implementation} \label{sec:implementation}
We implemented this mechanism on StreamFlow, a hybrid WMS based on the open standard Common Workflow Language (CWL). It has a well-defined module structure, allowing the community to develop some plugins to customise the default StreamFlow features, such as a new scheduling policy. 

The fault tolerance implementation has some features and optimisations that are not currently included in the semantics. 
% loop
A feature of StreamFlow that is supported in our implementation is the possibility of having loops in the workflow. Loops change nothing in our idea; indeed, the \textit{provenance graph} is always a Direct Acyclic Graph (DAG) because the iterations already executed are unfolded in the provenance. Furthermore, thanks to creating a new workflow, the loop management is responsible for the workflow engine, in this case, the StreamFlow driver, which already supports loops. 
However, it was necessary to introduce a mechanism to stop the recovery workflow and its loop immediately after the failed step was recovered.
An optimisation regarding the semantics is synchronising recovery workflows that want to roll back the same step concurrently. The implementation introduces a dependency between the steps across different recovery workflows. The driver manages this synchronisation.

\begin{figure}[t]
    \centering
     \includegraphics[width=.7\linewidth]{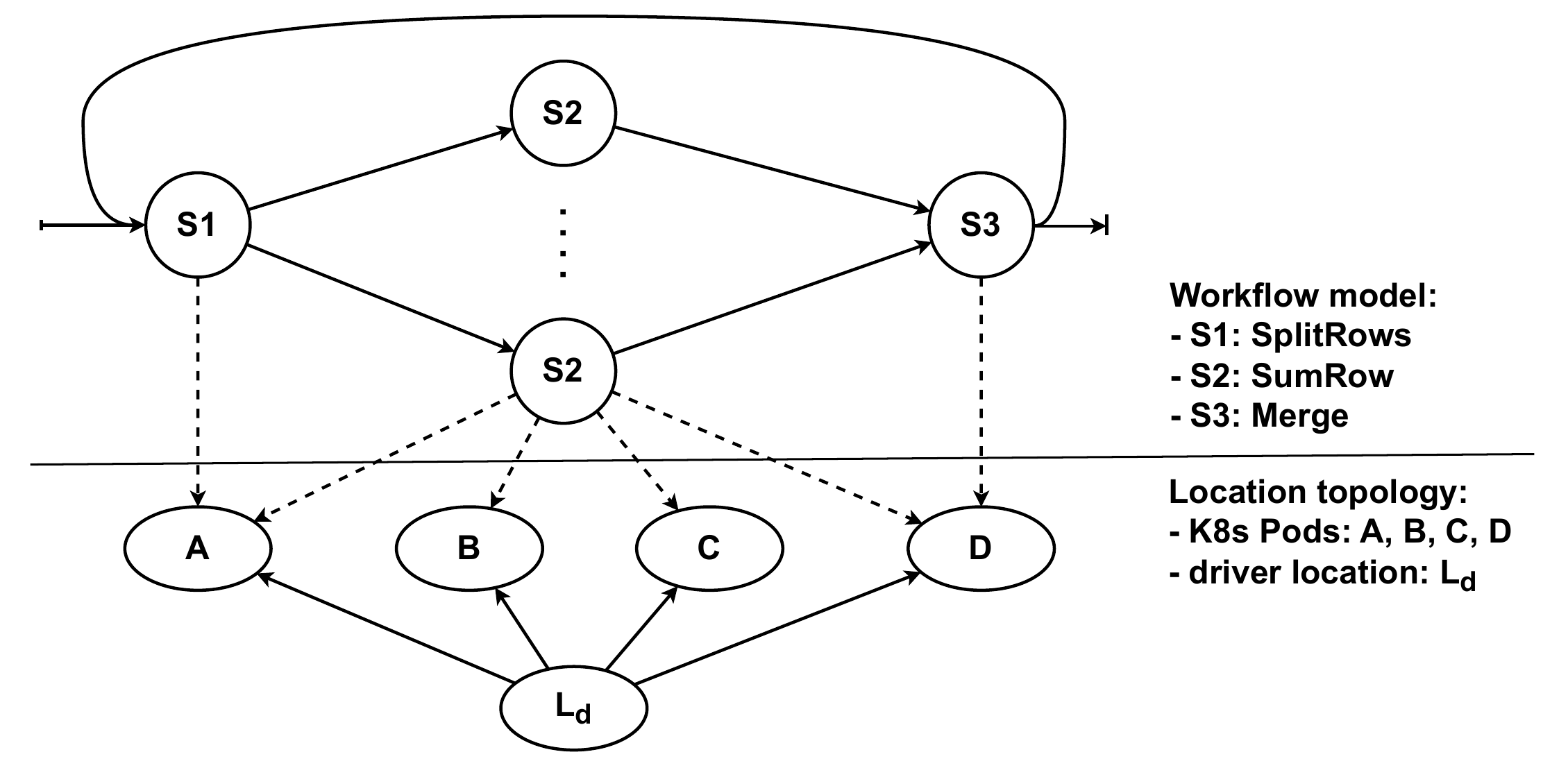}
    \caption{Workflow model presents a loop of 3 steps where the S2 (i.e. \textit{SumRow}) step has multiple instances. These steps are mapped on the A, B, C and D locations.}
    \label{fig:workflow-experiment}
\end{figure}

\subsection{Experiment environment}
We developed an application and encapsulated it in a Docker image\footnote{\url{ https://hub.docker.com/repository/docker/mul8/sf-failure/general}
}.
The application has some hyper-parameters to raise the failures. It is possible to choose the probability and the type of failure occurrence. The failure type can be \textit{tool}, i.e. soft error, or \textit{resource}, i.e. fail-stop error. 
The latter simulates a resource preemption or a shutdown of the location for any reason. The application forces the Docker container to terminate with error  to simulate the fail-stop error.

In the experiments, we ran the StreamFlow driver on our local machine, which also contains the dataset. Instead, we configured a virtualised Kubernetes~(K8s) cluster for the remote locations. The K8s cluster has 3 control plane nodes and 4 worker nodes. Each node has 4 virtual CPUs and 8GB of memory.
In our experiment, we used 4 Pods and each one uses our Docker image. When the Docker container exits with an error, it also fails the Pod, and K8s will restart it. All data in the Pod will be lost because it does not have a persistent volume.

\begin{figure}
    \vspace{-2mm}
    \centering
    \includegraphics[width=.7\linewidth, trim={6cm 2cm 2cm 0cm},clip]{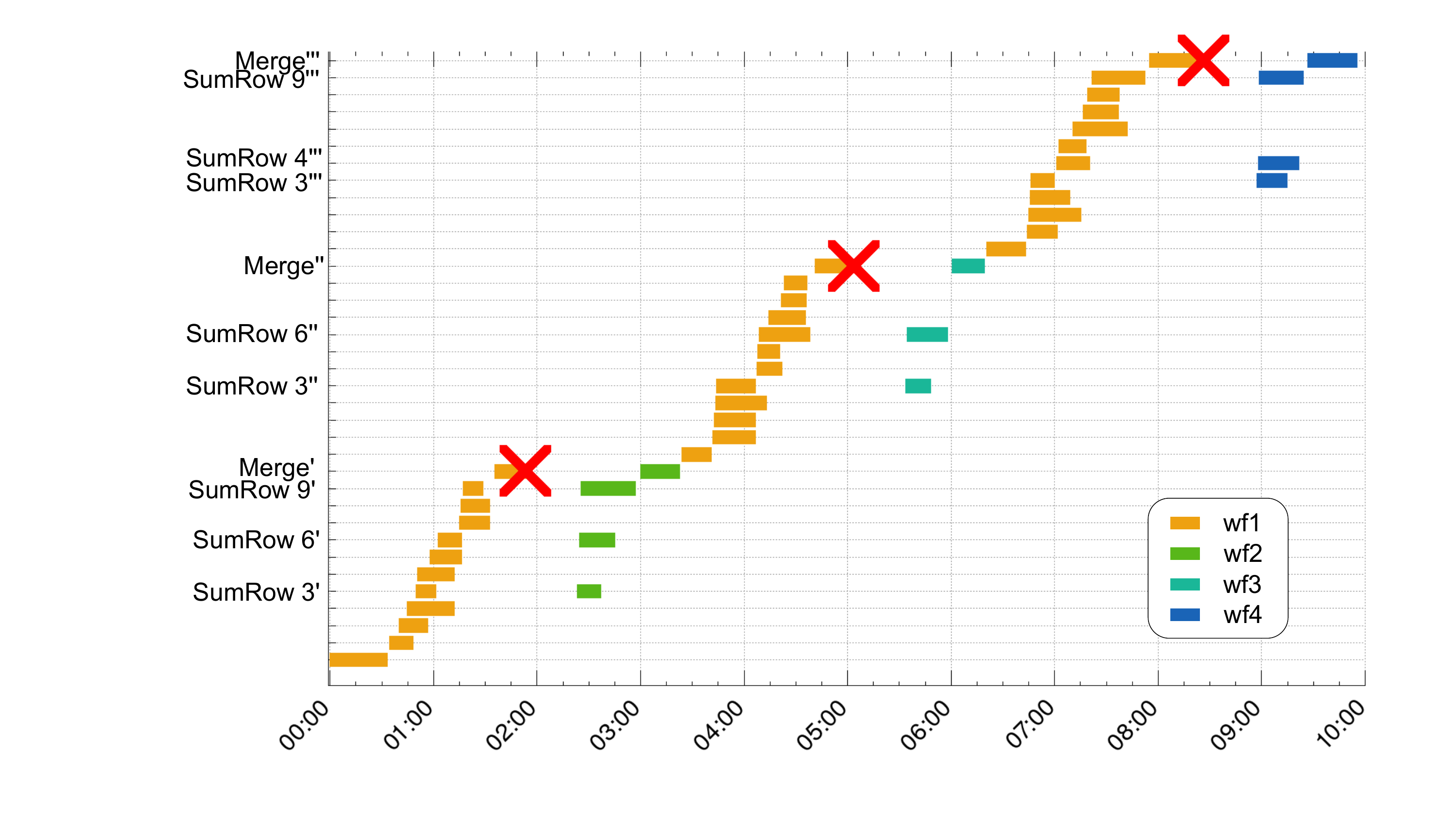}
    \caption{Execution of the workflow in Fig.~\ref{fig:workflow-experiment}. Some step names are omitted for the sake of readability. The execution is represented in minutes and seconds.}
    \label{fig:snd-experiment}
    \vspace{-2mm}
\end{figure}

\subsection{Experiment}

Fig.~\ref{fig:workflow-experiment} shows the workflow configuration. 
The steps are enclosed in a loop; the \textit{SumRow} step has different instances based on how much output data the \textit{SplitRows} generate.
The exit condition of the loop is to reach the i-th iteration. In this experiment, we set it at three iterations; moreover, the \textit{SplitRows} generates 10 data, so there will be 10 \textit{SumRow} step instances.
We used four locations, i.e. K8s Pods, named A, B, C and D. The \textit{SplitRow} step is mapped on the location A and the \textit{Merge} step is mapped on the location D. \textit{SumRow} instances are mapped on all the locations, and each is scheduled on the first available location. Moreover, we set the location failure only in the \textit{Merge} step with 50\% of probability.

\begin{figure}[t]
    \centering
    \includegraphics[width=.75\linewidth]{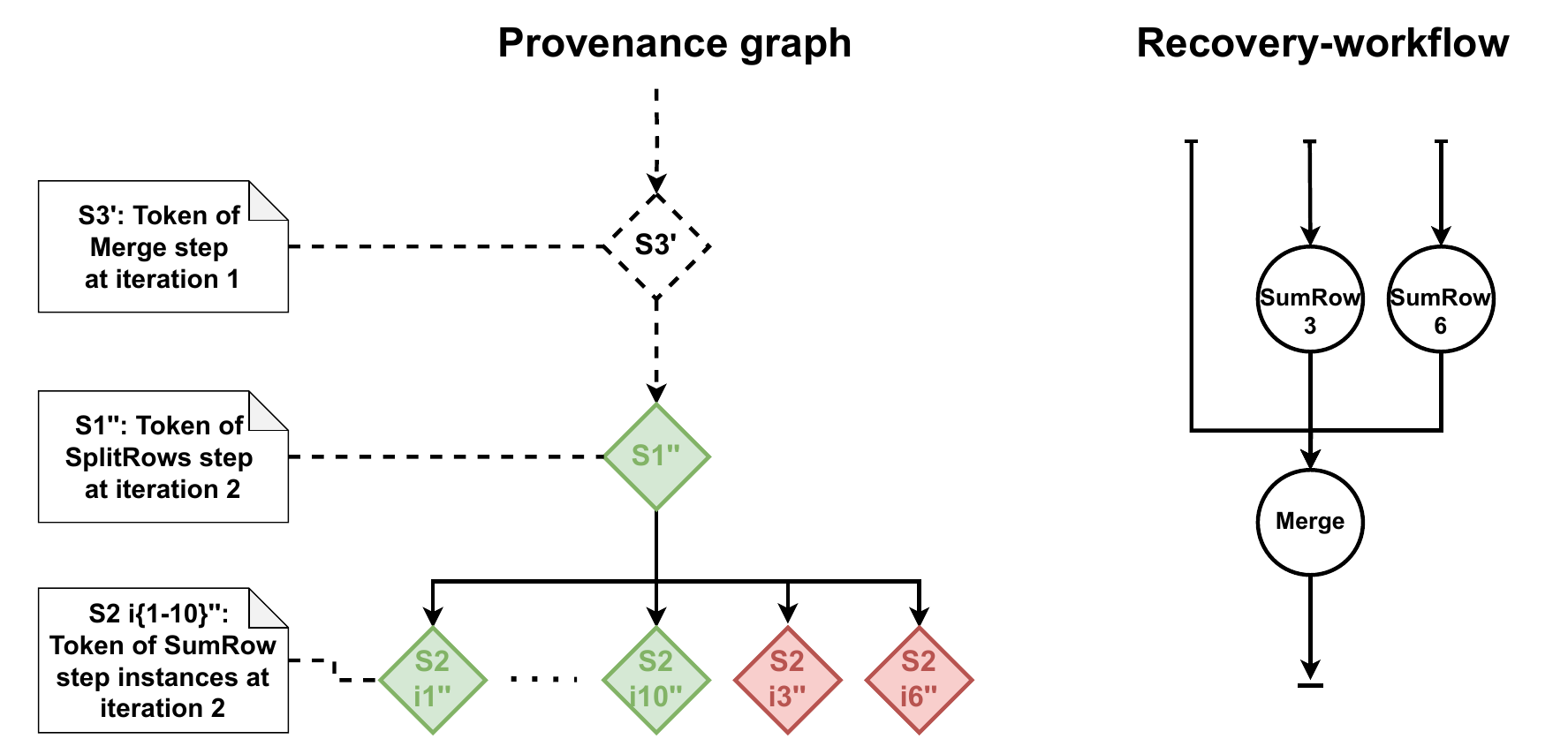}
    \caption{Left: the visited \textit{provenance graph}. Right: the created recovery workflow.}
    \label{fig:mechanism}
    \vspace{-5mm}
\end{figure}

Fig.~\ref{fig:snd-experiment} shows an example of execution. The bars represent the step execution, and the cross represents the failure of the step. The wf1 is the original workflow. Instead, the wf2-4 are recovery workflows. In this execution, the container in \textit{Merge} step throws the location failure at each iteration. 
The recovery mechanism creates a recovery workflow with the appropriate steps to rollback (e.g., the wf3, which is the teal-colored bars in Fig.~\ref{fig:snd-experiment}). 
For example, Fig.~\ref{fig:snd-experiment} shows the \textit{Merge} step fails at the second iteration, called \textit{Merge''}, while Fig.~\ref{fig:mechanism} shows the recovery mechanism of the failure. In the \textit{provenance graph}, the green diamonds are the available data in some locations, and the red diamonds are lost data.
The visit of the \textit{provenance graph} starts from its input data; in this case, there are only two lost data. In particular, these lost data are produced by \textit{SumRow 3''}  and \textit{6''}. We lose these data because these two steps have been scheduled and executed on the same location of \textit{Merge''}, i.e. location D, and the data were not copied to other locations.
The recovery workflow is created with 3 steps. The available data are in the input of the \textit{Merge} step. Meanwhile, the \textit{SumRow} 3'' and 6'' steps are re-executed. 
In this experiment, we do not use fully the potential of hybrid workflows because we used only homogeneous locations. However, the approach does not change. If a location fails but becomes available again, as in the case of K8s Pod, it can be used to re-execute the step. 
Otherwise, it is possible to apply a scheduling policy that changes the location to retry the step execution. In the implementation, there is a delay, choosen by the user, before retrying to communicate with a failed location; indeed, in the experiment, the delay was set to 20 seconds.
These 20 seconds of delay can be seen in Fig.~\ref{fig:snd-experiment}; in fact, there are idle times between the failure steps and the first step of the recovery workflows.

\section{Conclusions}\label{sec:conclusion}
Given the escalating computational demands of scientific applications, their execution time remains a significant challenge. Therefore, a robust system that safeguards against the loss of hours or days of computation is crucial. This underscores the importance of WMSs and their role in ensuring such applications' reliability.
Moreover, the hybrid workflow paradigm, a relatively new and versatile approach, has gained traction recently. In this work, we show a fault tolerance mechanism for hybrid workflows. We also model the idea with semantics, showing an example of execution. Then, we implement the concept and show a workflow with different patterns, such as loops and multi-instances.

One key advantage of our fault tolerance mechanism is the absence of complex logic in the workflow engine to restore previous internal states or undo some actions.
Instead, it is just necessary to synchronise different workflow executions. Finally, we can retrieve the successful steps that produce the workflow output by navigating the provenance graph, which is built across different workflows.
Future work involves evaluating implementation overhead, especially with more complex configurations using real-case workflows and different numbers of locations. It also involves combining the retry-rollback with checkpointing the data in safe locations. Other future work involves aligning semantics with the implementation describing synchronisation and loops and extending both with nondeterministic workflows.

\begin{credits}
\subsubsection{\ackname}
    This work was supported by: the Spoke 1 ``FutureHPC \& BigData'' of ICSC - Centro Nazionale di Ricerca in High-Performance Computing, Big Data and Quantum Computing, funded by European Union - NextGenerationEU; the EUPEX EU’s Horizon 2020 JTI-EuroHPC research and innovation programme project under grant agreement No 101033975.
\end{credits}

\clearpage

\appendix
\section{Appendix}\label{sec:appendix}
\paragraph{Explanation of the rules} 
Rule (\textsc{Init}) initialise the execution of the step on the location it is mapped to. With this rule the diver location adds the trace of the step $s$, $\trace(s)$ to its deploying location.

Rule (\textsc{Exec}) triggers the computation of the step producing or the resulting data end message $\msg{d,l}$ sent to the driver to confirm the creation of the data; or an error. Depending if the input data is present on the location or not, the error could be in the execution of the step, or because the input data is missing (probably due to the crash of the location).

There is a distinction between transferring a data and messages of the confirmation and transferring the error messages. The rule (\textsc{Trans}) is transferring a message/data from the location $l_1$ to the location $l_2$, where one of the locations could be a driver location. In the case of confirming messages: the data is produced, the data is received or message is received, when transferred, the message is removed from the set $\Data{l_1}$. When the action is done, the message is saved on the targeting location.

Transferring the error message with the rule (\textsc{TransErr}) would move the error message from the location which produces it to the driver location. In addition, this transition adds into the trace of the driver location the recovery predicate $\rec{x}$ which depends on the type of the error received. In that way the driver can perform the recovery of the failed step.

The rule (\textsc{LostData}) represents deleting the message of confirmation of production of data $d$ i.e. $\msg{d,l}$ in the presence of the message $\error{d,l}$ meaning that if the data $d$ is lost on the location $l$ and it was produced on the same location, it means that the step producing the data should be re-executed.

\begin{figure*}[t]
{\scriptsize
      \begin{mathpar}
        %%%%%%%%%%%%%%%%%%%%%%%%%%%%%%%%%%%%%%%%%%
         \inferrule*[left=(\textsc{Init})\;]
        { l=\mapping(s) }
        {\location{l_d}{\Data{l_d} }{\arancione{\pointer\init{\trace(s),l}.\trace_d}}\para
        \location{l}{\Data{l} }{\trace}
        \lts{} 
        \location{l_d}{\Data{l_d} }{\arancione{\init{\trace(s),l}.\pointer\trace_d}}\para
        \location{l}{\Data{l} }{\trace\arancione{\para \pointer\trace(s)}}
        }
        \and
        %%%%%%%%%%%%%%%%%%%%%%%%%%%%%%%%%%%%%%%%%%%%
         \inferrule*[left={\footnotesize{(\textsc{Exec})}}\;]
        {{\begin{cases} 
        \text{if } \;\In(s)\not\subset \Data{l}\;\wedge\; \msg{s}\in \Data{l},\;
        X=\{\error{D,l}\}\quad\text{where } D=\{d\para d\in \In(s)\;\wedge\; d\not\in \Data{l}\}\\
        \text{if } \;\In(s)\subset \Data{l}\; X=\{\Out(s),M_{d} \} \quad\text{where } 
        M_{d}=\{\msg{d,l}\para d\in \Out(s) \} \; \vee\; X=\{\error{s}\}
        \end{cases}} }
        {\location{l}{\Data{l} }{\arancione{\pointer\exec{s,\In,\Out}}.\trace}\lts{} 
        \location{l}{(\Data{l}\arancione{\setminus\{\msg{s}\} }) \arancione{\cup X} }{\exec{s,\In,\Out}.\arancione{\pointer}\trace} }
        \and
        %%%%%%%%%%%%%%%%%%%%%%%%%%%%%%%%%%%%
         \inferrule*[left=(\textsc{Trans})\;]
         {v\in \Data{l_1} \quad \wedge \quad {\begin{cases} 
         % \text{if } v=d, \text{ then } X=\emptyset \text{ and } V=\{d,\ok{d}\} \vee v'=\error{\msg{\rec{d,l_2}}}  \\
          \text{if } v\in\{d, \msg{s}\}, \text{ then } X=\emptyset \text{ and } V=\{v,\ok{v}\} \vee V=\{\error{v,l_2}\}  \\
        % \text{if } v=\msg{s}\text { \arancione{stesso sopra} }\quad \wedge \quad \forall d_i\in \In(s),\; \msg{d_i,l_1}\in\Data{l_1},  \text{ then } X=\emptyset \\
        \text{if }v= \{\msg{d,l_1},\ok{d},\ok{m_s}\} , \text{ then } X=\{v\}  \text{ and } V=\{v\}
        \end{cases}}}
         {\location{l_1}{\Data{l_1}}{\arancione{\pointer\transfer{v,l_2 }}.\trace_1} \para  \location{l_2}{\Data{l_2}}{\arancione{\pointer\transrec{v,l_1 }}.\trace_2} \lts{}\\
         \location{l_1}{\Data{l_1}\arancione{\setminus X}}{\transfer{v,l_2 }.\arancione{\pointer}\trace_1}\para \location{l_2}{\Data{l_2}\arancione{\cup V}}{\transrec{v,l_1 }.\arancione{\pointer}\trace_2}
         }
         \and
        % %%%%%%%%%%%%%%%%%%%%%%%%%
        % %%%%%%%%%%%%%%%%%%%%%%%%%
           \inferrule*[left=(\textsc{TransErr})\;]
         {v=\error{x} \quad \wedge \quad \error{x} \in \Data{l} \quad \text{where } x=\{s,(D,l), (d,l),(\msg{s},l)\}}
         {\location{l}{\Data{l}}{\arancione{\pointer\transfer{v,l_d }}.\trace} \para  \location{l_d}{\Data{l_d}}{\arancione{\pointer\transrec{v,l}}.\trace_d} \lts{}\\
         \location{l}{\Data{l}\arancione{\setminus\{v\}}}{\transfer{v,l_d }.\arancione{\pointer}\trace}\para \location{l_d}{\Data{l_d}\arancione{\cup \{v\}}}{\transrec{v,l}.\pointer\trace_d\arancione{\para \pointer\rec{x}}}
         }
         \and
         %%%%%%%%%%%%%%%%%%%%%%%%%%%%%%%%%%%%%%%%
         \and
         %%%%%%%%%%%%%%%%%%%%%%%%%
         \inferrule*[left=(\textsc{LostData})\;]
         {\error{d,l}\in D_{l_d} \quad \wedge\quad \msg{d_i,l}\in D_{l_d} }
         {D_{l_d}= D_{l_d}\setminus \{\msg{d_i,l}\} }
         %%%%%%%%%%%%%%%%%%%%%%%%%%%%%%
      \end{mathpar}}
      \caption{Reduction semantics rules.}
      \label{fig:semantics}
    \end{figure*}

The recovery actions are orchestrated by the driver location. There are four different recovery action that the driver can perform, depending on the type of error and the messages saved in its set $\Data{l_d}$: 

$(i)$ the recovery of a step by applying the rule (\textsc{Rec(s)}) used in the setting when the step failed its computation, but the necessary input data is present on the executing location. In this case the action $\rec{s}$ can be performed when $\error{s}\in\Data{l_d}$. To recover the step $s$, the driver is re-executing the trace corresponding to the step $s$, denoted by $\trace(s)$ and it is indicating to the executing location to re-execute the step $s$, buy adding the trace $\trace_{s}$ on it. In that way the recovery workflow is created to deal with the execution of the step $s$. When step is executed and the resulting data is obtained, the original trace $\trace$ can continue the execution.

$(ii)$ recovery of the data $d$ necessary to re-execute the step $s$ could have three options:

-the case when the data $d$ is present on the location where it was produced. In this case driver is adding the transfer action in the execution trace of the corresponding location; 

-the case when the data is not present on the location it produce it and it is necessary to execute the steps which produce data; and 

-when the location containing the missing data is exactly the driver location. In the final case, the transfer of the data is already belonging to the trace added to the driver location ($\trace_d(s)$).

Rules (\textsc{Rec(d)}) and (\textsc{Rec($\msg{s}$)}) allow for redoing of the transfer action. In this case the errors are given due to fault behaviour during the data transfer and it is necessary to redo just the transfer of the data.

\begin{figure*}[]
    {\scriptsize
      \begin{mathpar}
        \inferrule*[left=(\textsc{Rec(s)})\quad]
        {\error{s}\in\Data{l_d} }
        {\location{l_d}{\Data{l_d}}{\trace_d\para \arancione{\pointer\rec{s}}} \para 
        \location{l}{\Data{l}}{\trace(s)}\lts{}\\
        \location{l_d}{\Data{l_d}\arancione{\setminus\{\error{s}\}}}{\trace_d\para \rec{s}\arancione{\pointer\para \pointer\trace_d(s\setminus \trace_{d_{\In(s)}})}} \para 
        \location{l}{\Data{l}}{\nul\para \arancione{\trace_{s}}}}
         \and
        %%%%%%%%%%%%%%%%%%%%%%%%%%%%%
        %%%%%%%%%%%%%%%%%%%%%%%%%%%%%
        \inferrule*[left=(\textsc{Rec(D)})\quad]
        {\forall d_i\in D\text{ such that } \error{d_i,l},\msg{d_i,l_j}\in\Data{l_d}\\
        \forall d_h\in D\text{ such that } \error{d_h,l}\in\Data{l_d}\text{ and }\msg{d_h,l'},d_h\notin \Data{l_d}\and d_h\in \In(s_r)
         }
        {\location{l_d}{\Data{l_d}}{\trace_d\para \arancione{\pointer\rec{D,l}}} \para 
        \prod_{j}\location{l_j}{\Data{l_j}}{\trace_j}\para 
        \prod_{k}\location{l_k}{\Data{l_k}}{\trace_k}\para 
        \location{l}{\Data{l}}{\trace}
        \lts{}\\
        \location{l_d}{\Data{l_d}\arancione{\setminus\{\error{D,l}\}}}{\trace_d\para \rec{D,l}\arancione{\pointer} \arancione{\para \pointer\trace_d(s)\para\rec{s_r}}} \para \\
        \prod_{j}\location{l_j}{\Data{l_j}}{\trace_j\arancione{\para \pointer\transfer{d_i,l}}}\para 
        %\prod_{k}\location{l_k}{\Data{l_k}}{\trace_k}\para 
       \location{l}{\emptyset}{\nul\arancione{\para \trace(s)}}
        }
         \and
        %%%%%%%%%%%%%%%%%%%%%%%%%%%%%
        \inferrule*[left=(\textsc{Rec(d)})\quad]
        {\error{d,l},\msg{d,l'}\in\Data{l_d}
        %\quad \wedge\quad d\in\In(s)
        \quad \wedge\quad l'\in\mapping(s_1)\quad d\in\Out(s_1) }
        {\location{l_d}{\Data{l_d}}{\trace_d\para \arancione{\pointer\rec{d,l}}} \para 
        \location{l'}{\Data{l'}}{\trace'}\para 
        \location{l}{\Data{l}}{\trace}
        \lts{}\\
        \location{l_d}{\Data{l_d}\arancione{\setminus\{\error{d,l}\}}}{\trace_d\para \rec{d,l}\arancione{\pointer} } \para 
        \location{l'}{\Data{l'}}{\trace'\arancione{\para \pointer\transfer{d,l}}}\para \\
       \location{l}{\Data{l}}{\trace\arancione{\para \transrec{d,l'}.\conf{\ok{d},l}}}
        }
         \and
        %%%%%%%%%%%%%%%%%%%%%%%%%%%%%
            \inferrule*[left=(\textsc{Rec($\msg{s}$)})\quad]
        {\error{\msg{s},l}\in\Data{l_d}
        %\quad \wedge\quad d\in\In(s)
        %\quad \wedge\quad l'\in\mapping(s_1)\quad d\in\Out(s_1) 
        }
        {\location{l_d}{\Data{l_d}}{\trace_d\para \arancione{\pointer\rec{\msg{s},l}}} \para 
        \location{l}{\Data{l}}{\trace}
        \lts{}\\
        \location{l_d}{\Data{l_d}\arancione{\setminus\{\error{\msg{s},l}\}}}{\trace_d\para \rec{d,l}\arancione{\pointer} \transfer{\msg{s},l}.\confrec{\ok{\msg{s}},l}} \para \\
       \location{l}{\Data{l}}{\trace\arancione{\para \transrec{\msg{s},l_d}.\conf{\ok{m_s},l}}}
        }
         \and
        %%%%%%
      \end{mathpar}}
      \caption{Recovery semantics rules.}
      \label{fig:roll-semantics}
    \end{figure*}

\begin{example}\label{ex:workflow_full}
This example shows ho to model the hybrid workflow given in Fig.~\ref{fig:example} and 
its fault tolerance mechanism using the formal syntax and semantics rules given in Figs.~\ref{fig:semantics} and \ref{fig:roll-semantics}.

The initial state of the workflow in Fig.~\ref{fig:example} can be represented by using the syntax of Def.~\ref{def:syntax} in the following way:

{\footnotesize
\begin{align*}
\workflow = \location{l_d}{\Data{l_d}}{\pointer\trace_d, \trace_{\workflow}}
        \para \prod_{i=1}^{4} \location{l_i}{\emptyset}{\emptyset}
\end{align*}}

where $\Data{l_d}=\{\trace_{\workflow},\mapping(s_i),d,\msg{s_i}\}$, $i=1,\ldots,4$, $\trace_{\workflow}=\prod_{i=1}^{4} \trace(s_i)$ and:

{\footnotesize
\begin{align*}
\trace_d = &\init{\trace(s_1),l_1}.(\transfer{\msg{s_1},l_1}.\confrec{\ok{\msg{s_1}},l_1} \para \transfer{d,l_1}.\confrec{\ok{d},l_1} ).\\
&\qquad(\prod_{i=1}^{2} \transrec{\msg{d_i,l_1},l_1}+\transrec{\error{x},l_1} )\para\\
%s2
&\init{\trace(s_2),l_1}.(\transfer{\msg{s_2},l_1}. \confrec{\ok{\msg{s_2}},l_1}\para \\
&\qquad\confrec{\ok{d_1},l_1}).(\transrec{\msg{d_3,l_1},l_1}+\transrec{\error{x},l_1})\para\\
%s3
&\init{\trace(s_3),l_2}.(\transfer{\msg{s_3},l_2}. \confrec{\ok{\msg{s_3}},l_2}\para\\
&\qquad\confrec{\ok{d_2},l_2}). (\transrec{\msg{d_4,l_2},l_2}+\transrec{\error{x},l_2})\para\\
%s4
&\init{\trace(s_4),l_1}.(\transfer{\msg{s_4},l_1}.\confrec{\ok{\msg{s_4}},l_1}\para \confrec{\ok{d_3},l_1} \para \\
&\qquad\confrec{\ok{d_4},l_1} ). (\transrec{\msg{d_5,l_1},l_1}+\transrec{\error{x},l_1})\\
\trace(s_1)=& (\transrec{d,l_d}.\conf{\ok{d},l_1}\para \transrec{\msg{s_1},l_d}.\conf{\ok{\msg{s_1}},l_1}).\\
&\qquad\exec{s_1,\{d\},\{d_1,d_2\}}.(\prod_{i=1}^{2} \transfer{\msg{d_i,l_1},l_d}+\transfer{\error{x},l_d}).\\
% &(\transfer{\msg{d_1},l_1,l_d}+\transfer{\error{x},l_1,l_d}\para \transfer{\msg{d_2},l_1,l_d}+\transfer{\error{x},l_1,l_d}).\\
&\qquad(\transfer{d_1,l_1}\para\transfer{d_2,l_2} )\\
\trace(s_i)=& (\transrec{d_{i-1},l_1}.\conf{\ok{d_{i-1}},l_{i-1}}\para \transrec{\msg{s_{i}},l_d}.\conf{\ok{\msg{s_{i}}},l_{i-1}}).\\
&\qquad\exec{s_i,\{d_{i-1}\},\{d_{i+1}\}}.\\
&\qquad(\transfer{\msg{d_{i+1},l_{i-1}},l_d}+\transfer{\error{x},l_d}).\transfer{d_{i+1},l_1}\\
\trace(s_4)= & (\transrec{d_3,l_1}.\conf{\ok{d_3},l_1}\para \transrec{d_4,l_2}.\confrec{\ok{d_4},l_1}\para\transrec{\msg{s_4},l_d}.\conf{\msg{s_{4}},l_1}).\\
&\qquad\exec{s_4,\{d_3, d_4\},\{d_5\}}.
(\transfer{\msg{d_5,l_1},l_d}+\transfer{\error{x},l_d})
\end{align*}}

The following text shows how the semantics rules work using the example presented. To simplify the presentation and highlight the executing action, if the context of the executing action is not important for the execution, it is denoted with $\context{}$. For instance, when executing transfer of the message $\msg{s_1}$, instead of writing 

{\footnotesize
$$(\transrec{d,l_d}\para \arancione{\pointer\transrec{\msg{s_1},l_d})}.\exec{s_1,\{d\},\{d_1,d_2\}}$$}

we write $\context{\arancione{\pointer\transrec{\msg{s_1},l_d})}}$.

The workflow execution starts with $\init{}$ action:
{\footnotesize
\begin{align*}
&\location{l_d}{\Data{l_d}}{\context{\arancione{\pointer\init{\trace(s_1),l_1}}}\para \pointer\trace_{d(s_2)}\para \pointer\trace_{d(s_3)}\para \pointer\trace_{d(s_4)}}   \para 
\location{l_1}{\emptyset}{\nul}\arancione{\lts{}}\\
%%%%%%%
&\location{l_d}{\Data{l_d}}{\context{\init{\trace(s_1),l_1}\arancione{\pointer}}\para \pointer\trace_{d(s_2)}\para \pointer\trace_{d(s_3)}\para \pointer\trace_{d(s_4)}}   \para 
\location{l_1}{\emptyset}{\nul\arancione{\para \trace(s_1)}}
\end{align*}}

The transfer action transferring the data $d$ from the driver location to the location $l_1$. Only the parts of the workflow affected by the transition are shown:

{\footnotesize
\begin{align*}
&\location{l_d}{\Data{l_d}}{\context{\arancione{\pointer\transfer{d,l_1}}}\para \pointer\trace_{d(s_2)}\para \pointer\trace_{d(s_3)}\para \pointer\trace_{d(s_4)}}   \para \\
&\location{l_1}{\emptyset}{\context{\arancione{\pointer\transrec{d,l_d}}} \para\pointer\trace(s_2)\para \pointer\trace(s_4)}\arancione{\lts{}}\\
%%%%%%%
&\location{l_d}{\Data{l_d}}{\context{\transfer{d,l_1}\arancione{\pointer}}\para \pointer\trace_{d(s_2)}\para \pointer\trace_{d(s_3)}\para \pointer\trace_{d(s_4)}}   \para \\
&\location{l_1}{\arancione{\{d,\ok{d}\}}}{\context{\transrec{d,l_d}\arancione{\pointer}} \para\pointer\trace(s_2)\para \pointer\trace(s_4)}\\
\end{align*}}

It is noticeable that, after the transition, the set $\Data{l_1}$ is not empty anymore, but it contains the data $d$. Additionally, the pointer is moved to the end of the prefixes $\transfer{}$ and $\transrec{}$. 
To show the execution of the step $s_1$, the location $l_1$ is in the state given below, where initial data $d$ and trigger message $\msg{s_1}$ are transferred on the location $l_1$ and saved in set $\Data{l_1}$ and the confirmation of the reception is sent to the driver:

{\footnotesize
 \begin{align*}
&\location{l_1}{\{d,\msg{s_1}\}}{\context{\arancione{\pointer\exec{s_1,\{d\},\{d_1,d_2\}}}} \para\pointer\trace(s_2)\para \pointer\trace(s_4)}\arancione{\lts{}}\\
&\location{l_1}{\Data{l_1}}{\context{\exec{s_1,\{d\},\{d_1,d_2\}}.\arancione{\pointer}} \para\pointer\trace(s_2)\para \pointer\trace(s_4)}
\end{align*} }
where $\Data{l_1}=\{d,d_1,d_2,\msg{d_1,l_1},\msg{d_2,l_1}\}$.
The step is computed successfully, the message $\msg{s_1}$ is removed from the location and the new data $d_1,d_2$ and messages $\msg{d_1,l_1},\msg{d_2,l_1}$ are produced.
With these basic actions/operations, the hybrid workflow semantics are presented. In the following, it is shown how the failure of the step execution is modelled and how the driver recovers the failed part of the workflow (fault tolerance mechanism). For this purpose, it is considered that workflow is in the state in which the data $d_1$ and $d_2$ are transferred to the corresponding locations and the messages from the driver to execute the related steps are delivered, triggering the execution of the steps $s_2$ and $s_3$.

In the following, the failures of executing steps $s_2$ and $s_3$ are modelled. Step $s_3$ fails only the computation of the step, and the data necessary to re-execute the step remains on the location $l_2$ (soft error). Instead, step $s_2$ fails due to location failure, during which all data saved on the location $l_1$ is lost (fail-stop error). The failure of the step $s_3$ is modelled by producing the error message $\error{s_3}$:

{\footnotesize
\begin{align*}
    &\location{l_2}{\{d_2,\msg{s_3}\}}{\context{\arancione{\pointer\exec{s_3,\{d_2\},\{d_4\}}}}}\rosso{\lts{}}\\
    %%%%
&\location{l_2}{\{d_2,\rosso{\error{s_3}}\}}{\context{\exec{s_2,\{d_2\},\{d_4\}}.\rosso{\pointer}}}
\end{align*}}

The error is communicated to the driver location with:

{\footnotesize
\begin{align*}
    &\location{l_2}{\{d_2,\rosso{\error{s_3}}\}}{\context{\rosso{\pointer\transfer{\error{s_3},l_d}}}}\para\\
    &\location{l_d}{\Data{l_d}}{\context{\rosso{\pointer\transrec{\error{x},l_2}}}}\rosso{\lts{}}\\
%%%%%%%
   &\location{l_2}{\{d_2\}}{\context{\transfer{\error{x},l_d}).\rosso{\pointer}}}\para\\
    &\location{l_d}{\Data{l_d}\cup \rosso{\error{s_3}}}{
\context{\transrec{\error{x},l_2}.\rosso{\pointer}}\rosso{\para \pointer\rec{s_3}}}
\end{align*}}

In this moment, the driver is able to coordinate the recovery of the step $s_3$ done by re-sending the message to trigger the re-execution of the step and by inserting into the trace of the location $l_2$ the sub-workflow to compute the part of the workflow that failed:

{\footnotesize
\begin{align*}
    &\location{l_{d}}{\Data{l_{d}}\cup \verde{\error{s_{3}}}}{\trace_{d(s_{1})}\pointer\para 
\trace'_{d(s_{2})}\para
\trace_{d(s_{2})}\pointer\para\pointer\trace_{d(s_{2})}\para\verde{ \pointer\rec{s_{3}}}}\para\\
&\location{l_{2}}{\{d_{2}\}}{\trace(s_{3_{\setminus Out}}).\arancione{\pointer}\transfer{d_{3},l_{1}}}\verde{\lts{}}\\
&\location{l_{d}}{\Data{d}}{\trace_{d(s_{1})}\pointer\para 
\trace'_{d(s_{2})}\para
\trace_{d(s_{2})}\pointer\para\pointer\trace_{d(s_{2})}\para\rec{s_3}\verde{ \pointer\para \pointer\trace_d({s_{3}}_{\setminus \trace_{\In(s_3)}})}}\para\\
&\location{l_{2}}{\{d_{2}\}}{\verde{ \nul\para \pointer\trace({s_{3}}_{\setminus \trace_{\In(s_3)}}) }}
\end{align*}}
The traces $\pointer\trace_d({s_{3}}_{\setminus \trace_{\In(s_3)}})$ and $\pointer\trace({s_{3}}_{\setminus \trace_{\In(s_3)}})$ represent the recovery workflow (recovery actions) to be executed to retrieve the step $s_3$ and the result of its execution. 
The model of the failure of the step $s_2$ is presented below. The location failed, and all data are lost, which produces the error of the data $d_1$ (so the message $\error{d_1,l_1}$). In semantics:

{\footnotesize
\begin{align*}
    &\location{l_1}{\emptyset}{\trace(s_1)\pointer\para\context{\rosso{\pointer\exec{s_2,\{d_1\},\{d_3\}}}} \para \pointer\trace(s_4)}\rosso{\lts{}}\\
    %%%%%
    &\location{l_1}{\{\rosso{\error{d_1,l_1}}\}}{\trace(s_1)\pointer\para\context{\exec{s_2,\{d_1\},\{d_3\}}.\rosso{\pointer}} \para \pointer\trace(s_4)}
\end{align*}}

After the message is produced, it needs to be communicated to the driver location:

{\footnotesize
\begin{align*}
    &\location{l_1}{\{\rosso{\error{d_1,l_1}}\}}{\context{\rosso{\pointer\transfer{\error{d_1,l_1},l_d}}}}\para\\
    &\location{l_d}{\Data{l_d}}{
\context{\rosso{\pointer\transrec{\error{x},l_1}}}}\rosso{\lts{}}\\
%%%
&\location{l_1}{\emptyset}{\context{\transfer{\error{d_1,l_1},l_d}.\rosso{\pointer}} }\para\\
    &\location{l_d}{\Data{l_d}\cup \rosso{\error{d_1,l_1}}}{
\context{\transrec{\error{x},l_1}.\rosso{\pointer}}\rosso{\para\pointer\rec{d_1,l_1}}}
\end{align*}}

Having on the driver location the error message about the crash of the location $l_1$ in which data $d_1$ is lost, it is necessary to remove the previous message confirming the data $d_1$ produced on the $l_1$. In that way, the driver knows that the step which produce data $d_1$ and $d_2$ needs to be re-executed, therefore:

{\footnotesize
\[
\{d,\msg{d_2,l_1},\msg{d_1,l_1},\error{d_1,l_1}\}\lts{}\{d,\error{d_1,l_1}\}
\]}

Now, since $\error{d_1,l_1}\in \Data{d}$ and $\msg{d_1,*}\notin \Data{d}$, meaning that the data $d_1$ is not present on the location which produced it, it is necessary to recover the step producing it:

{\footnotesize
\begin{align*}
    &\location{l_d}{D'_{l_d}}{\context{\verde{\pointer\rec{d_1,l_1}}}}\verde{\lts{}}\\
    &\location{l_d}{D'_{l_d}}{\context{\rec{d_1,l_1}\verde{\pointer}}\verde{\para \pointer\rec{s_1}\para \pointer\trace_d(s_1)}}
\end{align*}}

The recovery of the step $s_1$ implies that the location $l_1$ needs to execute the step $s_1$ one more time. Since the location $l_1$ crashed, the data necessary to execute the step is not present; therefore, the execution of $s_1$ produces an error $\error{d,l}$ meaning that the data $d$ is missing. In that case, the driver receiving the error message and having the data $d$ will transfer the data directly to the location $l_1$.

\end{example}

\end{document}